\documentclass[%
 reprint,
 amsmath,amssymb,
 aps,
]{revtex4-1}

\usepackage{graphicx}
\usepackage{dcolumn}
\usepackage{bm}
\usepackage{verbatim}
\usepackage{array}
\usepackage{amsmath}
\usepackage{lineno}
\usepackage[bookmarksnumbered, pdfstartview=FitH,colorlinks,urlcolor=blue, citecolor=blue,linkcolor=blue,]{hyperref}

\renewcommand{\arraystretch}{0.8}

\begin{document}

\preprint{APS/123-QED}

\title{Search for Majoron at the COMET Experiment}

\author{Tianyu Xing$^{1,2}$}
\email{xingty@ihep.ac.cn}
\author{Chen Wu$^{3}$}
\email{wuchen1106@gmail.com}
\author{Han Miao$^{1,2}$}
\email{miaohan@ihep.ac.cn}
\author{Haibo Li$^{1,2}$}
\author{Weiguo Li$^{1,2}$}
\author{Ye Yuan$^{1,2}$}
\author{Yao Zhang$^{1}$}
\affiliation{%
	$^{1}$Institute of High Energy Physics, Chinese Academy of Sciences, Beijing 100049, China\\
	$^{2}$University of Chinese Academy of Sciences, Beijing 100049, China\\
	$^{3}$Research Center for Nuclear Physics, Osaka University, Ibaraki, Osaka 567-0047, Japan
}%


\date{\today}

\begin{abstract}
\begin{description}
\item[Abstract]
A new Goldstone particle named Majoron is introduced in order to explain the origin of neutrino mass by some new physics models assuming that neutrinos are Majorana particle. By expanding signal region and using likelihood analysis, it becomes possible to search for Majoron at experiments that is originally designed to search for $\mu-e$ conversion. For the COMET experiment, the sensitivity of process $\mu \rightarrow eJ$ is able to reach ${\mathcal{B}}(\mu \rightarrow eJ)=2.3\times 10^{-5}$ in Phase-I and $O(10^{-8})$ in Phase-II. Meanwhile, the sensitivities to search for Majoron at future experiments are also discussed in this article.
\item[Keywords]
Majoron, neutrino, COMET, new physics
\end{description}
\end{abstract}

\maketitle

\section{Introduction}

With the observation of neutrino oscillations, it is widely accepted that at least two neutrinos have mass~\cite{Schwetz_2011,Schwetz_2011_2}. There are two distinct theoretical models to explain the origin of neutrino mass: (1) neutrinos are Dirac particles, in which case they possess a lepton number conserving mass term; (2) neutrinos are Majorana particles and their mass term violates lepton number~\cite{MAALAMPI1980217,Mohapatra}. In the latter case, lepton number will inevitably be broken once assuming neutrinos are massive Majorana particle~\cite{CHIKASHIGE1981265}. If lepton number is a spontaneously-broken global symmetry, a massless Goldstone boson, Majoron ($J$) appears in theory as an unavoidable consequence of the spontaneous breaking of the lepton number.

The width of Z boson was once a good examination for models raised in early theoretical researches about Majoron. According to these models a large additional contributions to the invisible decay width of Z boson were predicted~\cite{GELMINI1981411,AULAKH1982136}. The invisible decay was observed after the Z-width measurement at LEP~\cite{DECAMP1990399,AARNIO1989539,AKRAWY1989530,ADEVA1989509}, but the measurement is consistent with the Standard Model prediction for the Z decay into three generations of neutrinos, which excludes the possibility of a large contribution due to Majoron~\cite{GONZALEZGARCIA1989383,ROMAO1990371}. Thereafter, new models have been developed by taking spontaneous breaking of R-parity into consideration, where there is no significant invisible Z-width~\cite{MASIERO1990273} and the decay of charged lepton with Majoron emission is permitted. Theoretical calculation implies that the branching fraction of $\mu \to eJ$ is potentially able to be measured in current experiments~\cite{ROMAO1991369,Hirsch}. The phenomenology of Majoron was discussed in many recent papers and so were, in particular, its lepton-flavour-violating (LFV) interactions, which provide strong supports in theory to the search for Majoron~\cite{Escribano_2022,Brune,ChengYu}.

Recently the models predicting the flavour-violating muon decay $\mu \to e X$ are revisited and have caused wide concern all over the world, where $X$ is an invisible boson with a mass smaller than the muon mass $m_{\mu} = 105.658~{\rm MeV}/c^{2}$~\cite{HEECK2018385,Heeck_2017wgr,Calibbi,Davidson}. There are some candidates for new particle $X$, such as light gauge bosons~\cite{IBARRA2022136933} and light (pseudo-)scalars~\cite{HEECK2018385} including familon~\cite{BABU1986360}, Majoron or axion-like particle (ALP)~\cite{Calibbi2021looking}. $\mu \to e X$ will be a good probe for experimentally investigating and verifying all these new particles predicted by LFV physics.

\subsection{Searching for Majoron with free $\mu$}

Several experiments have been performed to search for the inclusive two-body decay of positive muon $\mu^+ \to e^+ X$, where $X$ is a neutral boson, by precise measurement of energy spectrum of $e^+$.

In 1986, using $1.8 \times 10^7$ polarized $\mu^+$, the upper limit of branching fraction is determined to be ${\mathcal{B}}(\mu^+ \to e^+ X)<2.6 \times 10^{-6}$ at the 90\% confidence level (C. L.)~\cite{jodidio1986search}. However, in this research the $X$ was considered to be massless and the momentum distribution of signal $e^{+}$ was considered to be symmetric with the assumption that the signal $e^+$ are emitted isotropically. This assumption is not valid for the case of Majoron that couples to a V-A leptonic current, resulting in the fact that the signal is not isotropical for a polarized $\mu^+$ beam. Actually this research was not sensitive to $\mu^+ \to e^+ J$ because $e^+$ was detected in the direction opposite to the $\mu^{+}$ polarization where the observable, the endpoint of the Michel spectrum, drops to zero, so that the result is not valid for Majoron.

About 30 years later, the limit on the $\mu \to e X$ decay is measured again to be ${\mathcal{B}}(\mu^+ \to e^+ X)<5.8 \times 10^{-5}$ at the 90\% C. L. by TWIST experiment~\cite{TWIST} using a dataset of $5.8 \times 10^{8}$ events with assuming $X$ to be a light boson (with the mass of $X$ $m_{X} < 13~{\rm MeV}/c^{2}$) that couples to a V-A leptonic current. This is indeed the case for Majoron models in which the $e^+$ emission features the same asymmetry as the standard Michel decay so the result is the current best limit on $\mu \to e J$ decay. Branching ratio limits of order $O(10^{-5})$ are also obtained with different decay asymmetries for $X$ with masses $13~{\rm MeV}/c^{2}<m_X<80~{\rm MeV}/c^{2}$.

To search for the $X$ more precisely, Mu3e experiment will reach a sensitivity of ${\mathcal{B}}(\mu^+ \to e^+ X) \sim O(10^{-8})$ within the mass region $35~{\rm MeV}/c^{2}<m_X<95~{\rm MeV}/c^{2}$ in the near future, while there is a deterioration of the sensitivity within the mass region of $m_X<35~{\rm MeV}/c^{2}$ by about one order of magnitude (thus up to $\sim O(10^{-7}$) due to the difficulty of calibration~\cite{Perrevoort}.

\subsection{Searching for Majoron with muonic atoms}
Different from the monoenergetic signal in the decay of a free $\mu^+$, the electron energy spectrum in the decay of a $\mu^-$ in orbit has a width because of the nuclear recoil, of which the maximum energy is close to the signal energy of the $\mu - e$ conversion (the neutrinoless, coherent transition of a muon to an electron in the field of a nucleus, $\mu^{-} N \to e^{-} N$)~\cite{COMETTDR}. This fact introduces the possibility and advantages to search for Majoron with muonic atoms by using the detector of the $\mu - e$ conversion experiments within a preferable energy region with a large signal-to-background ratio. Moreover, the shape and the nuclear dependence of the electron spectrum could provide substantial information of the new physics\cite{Uesaka}. 

One previous study focuses on searching for Majoron with muonic atoms formed by muon capture instead of free $\mu^+$\cite{Czarnecki}, in which some $\mu - e$ conversion experiments such as COMET~\cite{COMETTDR} and Mu2e~\cite{Mu2eTDR} are taken into consideration. On one hand, with more intense beam available, the sensitivity will definitely be improved due to more $\mu^-$ captured. On the other hand, according to calculation in Ref.~\cite{Czarnecki}, the cross section of $\mu \to e X$ will get relatively larger compared with that of decay in orbit (DIO) $\mu^{-} \to e^{-} \nu_{\mu} \bar{\nu_{e}}$ with the increasing energy of detected electrons though the absolute cross section will drop. These features provide us a unique opportunity to search for Majoron in $\mu-e$ conversion experiments with at least the same sensitivity as former experiments using free $\mu^+$ according to the rough calculation in Ref.~\cite{Czarnecki}.

Since free $\mu^-$ are not related with the topic in this work, the following discussions are based on the orbital $\mu^-$ in the muonic atoms.

\section{Method of searching for Majoron on the COMET experiment}

\begin{figure}[ht]
	\centering
	\includegraphics[width=0.47\textwidth]{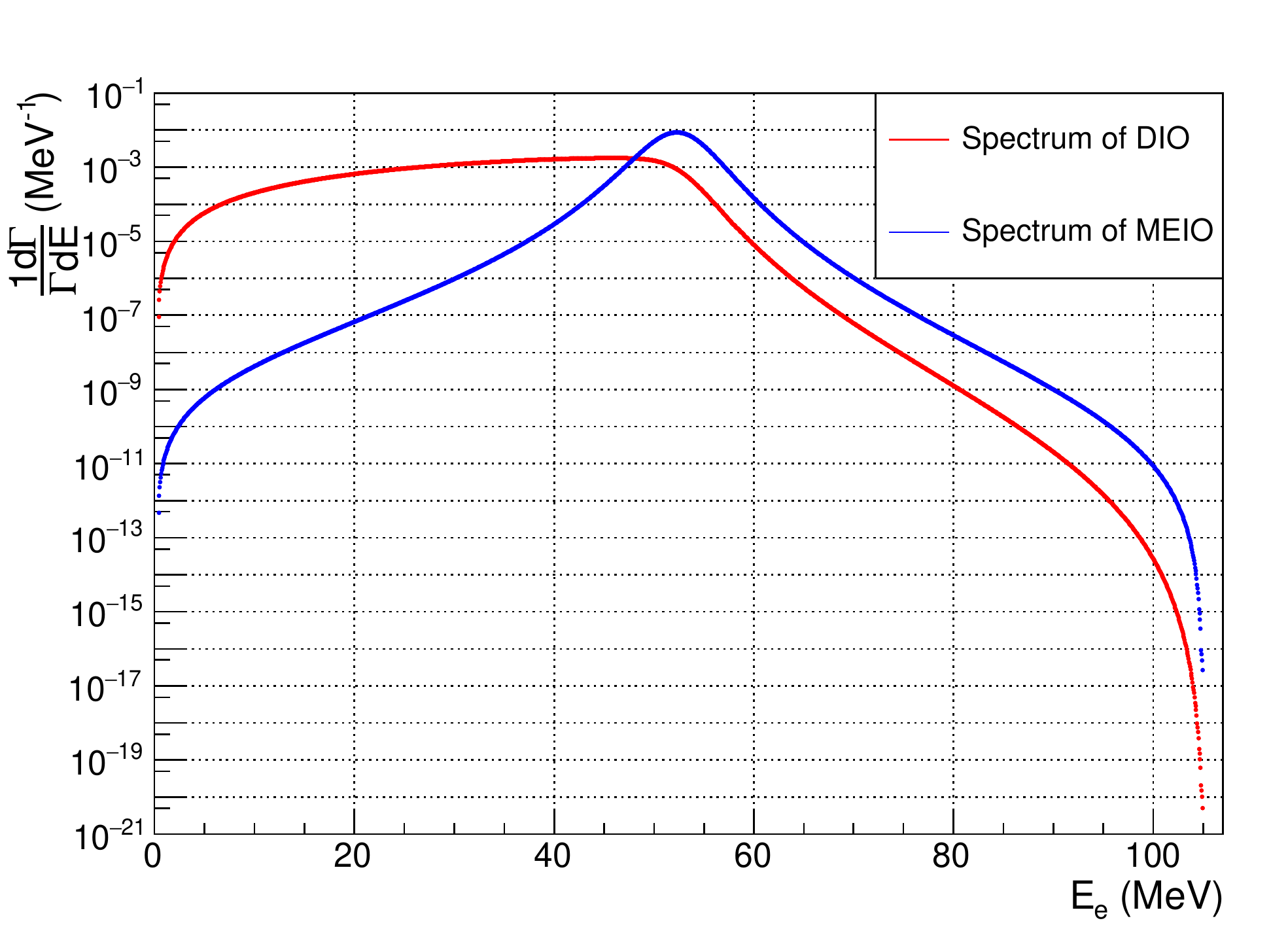}
	\caption{Spectra of $\mu^{-} \to e^{-} \nu_{\mu} \bar{\nu_{e}}$ (red) and $\mu \to e J$ (blue)}
	\label{fig:spectra}
\end{figure}

In this work, the target process is Majoron emission in orbit (MEIO) $\mu \to eJ$, while the dominant background process is decay in orbit (DIO). The energy spectra of the electrons from these two processes have been calculated by theoretical works~\cite{Czarnecki, Uesaka}, which are shown in Figure~\ref{fig:spectra}. The difference between the two energy spectra will be used to seek Majoron events. Utilizing about $10^{18}$ muon events taken by COMET Phase-I, energy spectra of electrons will be measured precisely. To improve the sensitivities, the distribution of detector acceptance is of great significance to get an accurate spectrum, which will be evaluated in Chapter~\ref{chap:3b}.

To extract the difference between the spectra, likelihood analysis method is used in this work.

\subsection{Likelihood analysis method}
Likelihood analysis is a very important and popular method of data analysis in particle physics. For an observable influenced by a host of involved parameters that include constants of nature in putative or speculative laws of physics as $\theta$, the probability to get a specific collection of measurements x of the observable can be expressed as $p(x\mid\theta)$ or $p(x;\theta)$. However in the actual experiments, usually we need to estimate parameters $\theta$ after obtaining the collection of observations $x$. In this case we evaluate the expression $p(x\mid\theta)$ only for the specific $x$ that was observed, and then examine how it varies as $\theta$ is changed. By constructing a likelihood function $\mathcal{L}(\theta)$, often also denoted as $\mathcal{L}(x \mid \theta)$ and sometimes $\mathcal{L}(\theta \mid x)$: $\mathcal{L}(\theta) = p(\theta \mid x) = p(x\mid\theta)$, which is evaluated at the given observed $x$ and considered to be a function of $\theta$, we can statistically describe how well the models with parameters fit the observations. This is what we called the likelihood method~\cite{cousins2020likelihood}.

In this work, the energy spectrum of electrons is expected to be the sum of the spectra of MEIO and DIO processes with a ratio $r$ that can be expressed as a formula $f(E)_{{\rm excepted}} = f(E)_{{\rm DIO}} + r f(E)_{{\rm MEIO}}$, where $f(E)_{{\rm excepted}}$ is the expected spectrum, $f(E)_{{\rm DIO}}$ and $f(E)_{{\rm MEIO}}$ denote the spectrum of DIO and MEIO process respectively. By comparing with the measured spectrum $f(E)_{{\rm measured}}$ using likelihood method, the ratio $r$ can be determined.

Since the COMET experiment is still waiting for taking data, simulation results are used in this work to predict the possibility of searching for Majoron.

\subsection{Extending the signal window}
For the COMET experiment, the branching fraction of $\mu \to eJ$ is determined by
\begin{equation}\label{eq:1}
{\mathcal B}(\mu \to e J) = \frac{N_{J}/(Acc_{J}f_{J})}{N_{{\rm \mu}}}\ ,
\end{equation}
where $N_J$ is the number of signal events, $N_{\rm \mu}$ is the total number of decayed orbital muons, $f_{J}$ is the fraction of $\mu \to e J$ within the available phase space constrained by the COMET detectors, and $Acc_{J}$ is the acceptance of the signal events.

In the case where Majoron is not observed, $N_J$ should be less than the statistical fluctuation of the total number of the electrons that could be detected $N_{\rm e}$. With assuming $N_{e}$ follows normal distribution, we have:
\begin{equation}
  \label{equ:upper_limit}
  N_J < 1.645\sigma_{e}~(90\%{\rm ~C.~L.}),
\end{equation}
where $\sigma_{\rm e}$ denotes the standard deviation of $N_{e}$. Considering that all of the detected electrons are from decayed orbital muons with ignorance of beam background, $N_{e}$ can be described by: 
\begin{equation}
  \label{equ:N_e}
  N_{e} = N_{\mu}Acc_{\rm dec}f_{\rm dec},
\end{equation}
where $f_{\rm dec}$ is the fraction of electrons within the available phase space constrained by the COMET detectors and $Acc_{\rm dec}$ is the acceptance of electrons in signal region. Then $\sigma_e$ will be: 
\begin{equation}
  \label{equ:sigma_e}
  \sigma_{e} = \sqrt{N_{\mu}Acc_{\rm dec}f_{\rm dec}}.
\end{equation}

Combining Eq.~(\ref{eq:1}), (\ref{equ:upper_limit}) and (\ref{equ:sigma_e}), we have: 
\begin{equation}\label{eq:2}
{\mathcal B}(\mu \to e J) < 1.645 \times \sqrt{ \frac{1} {N_{\mu} Acc_{\rm dec} f_{\rm dec}} } \frac{Acc_{\rm dec} f_{\rm dec}} {Acc_{J} f_{J}}\ ,
\end{equation}
which is proportional to two terms:
\begin{enumerate}
	\item statistics factor: term $\sqrt{\frac{1}{N_{\mu}Acc_{\rm dec}f_{\rm dec}}}$ is limited by the trigger rate which results from beam structure and tolerance of the detector;
	\item significance factor: term $\frac{Acc_{\rm dec}f_{\rm dec}}{Acc_{J}f_{J}}$ is decided by energy spectra and the bias of detector acceptance.
\end{enumerate}
A muonic atom will normally either decay in orbit or be captured by the nucleus $\mu^{-} N(A,Z) \to \nu_{\mu} N(A,Z-1)$. For the $\mu - e$ conversion experiments, the total number of decayed orbital muons is determined by
\begin{equation}
\begin{aligned}
N_{\mu} = \frac{1}{S.E.S._{\mu e}Acc_{\mu e}}\frac{\Gamma_{\rm decay}}{\Gamma_{\rm capture}}\ ,
\label{eq:muon_number}
\end{aligned}
\end{equation}
where $S.E.S._{\mu e}$ is single-event sensitivity of the $\mu - e$ conversion experiment, $Acc_{\mu e}$ is the acceptance of $\mu - e$ conversion events that are electrons with $104.971~{\rm MeV}$ energy, $\Gamma_{{\rm decay}}$ is the ratio of decayed muons over total orbital muons, and $\Gamma_{{\rm capture}}$ is the ratio of muons captured by nucleus over total orbital muons. Considering that only electrons will be recorded for no matter MEIO or DIO, the acceptance of the two processes are the same, namely $Acc_{\rm dec} = Acc_{J} = Acc_{e}$, where $Acc_{e}$ means the total acceptance of electrons within the signal range, so that the branching fraction can be finally determined by
\begin{equation}
\begin{aligned}
{\mathcal B}(\mu \to e J)
& < 1.645 \times \sqrt{\frac{\Gamma_{\rm capture} Acc_{\mu e} S.E.S._{\mu e}}{\Gamma_{\rm decay}}} \frac{1}{S}\ , \\
S &= \sqrt{\frac{Acc_{e} f_{J}^{2}}{f_{\rm dec}}}\ ,
\label{eq:3}
\end{aligned}
\end{equation}
where $S$ is significance factor that can be optimized to the maximum by changing the signal window.

\begin{figure}[h]
	\centering
	\includegraphics[width=0.47\textwidth]{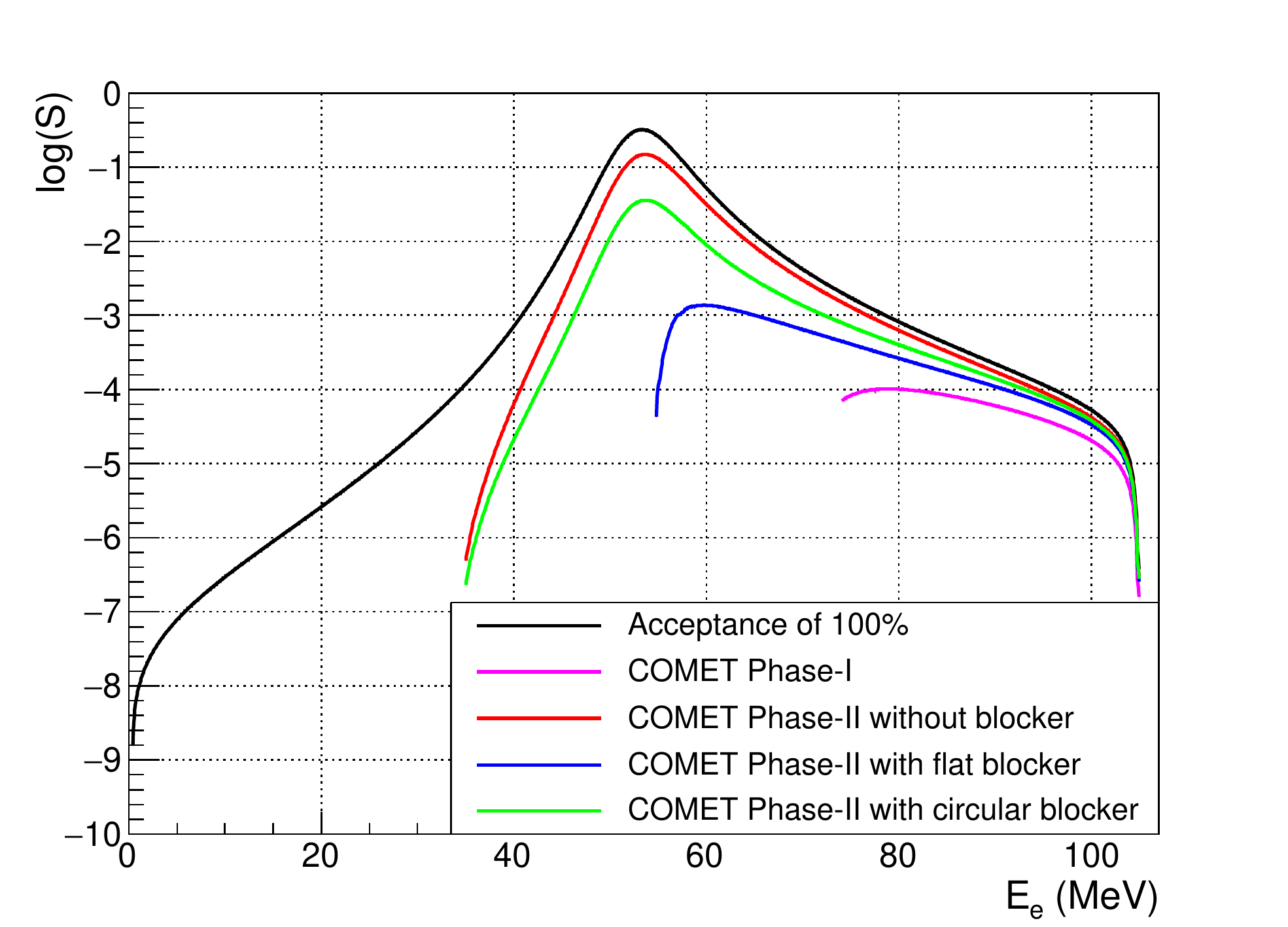}
	\caption{Distributions of the significance factor when varying lower limit of the signal window. The blocker is implemented in COMET Phase-II to control the hit rate in the detector, which will be explained for detail in Chapter~\ref{chap:4b}}
	\label{fig:significance}
\end{figure}

With the upper limit of signal window fixed at $105~{\rm MeV}$, a distribution of $S$ can be got by varying the lower limit. Several cases of COMET Phase-I and Phase-II are taken into consideration together with the ideal situation where the acceptance is 100\%. It is shown in Figure~\ref{fig:significance} that $S$ will gets maximum when the lower limit is around $80~{\rm MeV}$ in the case of COMET Phase-I. 

Therefore, the sensitivity to search for Majoron can be improved significantly by extending the signal window.

\section{Prediction of searching for Majoron on COMET Phase-I}

The COherent Muon to Electron Transition (COMET) experiment~\cite{COMETTDR} at the Japan Proton Accelerator Research Complex (J-PARC) in Tokai, Japan, aims to search for the neutrinoless coherent transition of a muon to an electron ($\mu - e$ conversion) in the field of a nucleus, with its single event sensitivity of $2 \times 10^{-17}$, which is more than four orders of magnitude improvement over the current upper limit of $7 \times 10^{-13}$ at 90\% C.L. from the SINDRUM II experiment~\cite{SINDRUMII_2006dvw}.

\subsection{Design of COMET Phase-I}
\label{chap:3a}

\begin{figure}[ht]
	\centering
	\includegraphics[scale=0.5]{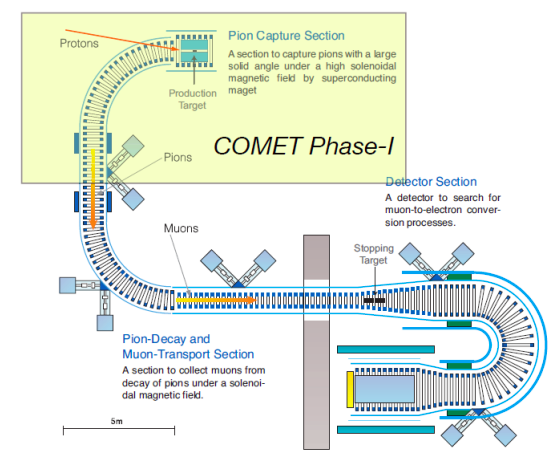}
	\caption{A schematic layout of COMET Phase-I (yellow box) and COMET Phase-II (whole figure)}
	\label{fig:twostage}
\end{figure}

To control potential backgrounds, the COMET experiment will be better optimized by precisely measuring muon beam. Therefore, a two-stage approach is adopted~\cite{COMETTDR, proceeding}. Figure~\ref{fig:twostage} shows the facilities of COMET Phase-I and Phase-II. A solenoid with $90^{\circ}$ bend is used to transport pions, which will decay into muons along the beam line. A hollow, cylindrical detector system (CyDet) surrounding the muon stopping target is specially designed and placed at the exit of the solenoid. Both the muon beam and the detector system are different between Phase-I and Phase-II, leading to the different acceptance of signals.

\subsection{Simulation of COMET Phase-I}
\label{chap:3b}

\begin{figure}[ht]
	\centering
	\includegraphics[width=0.45\textwidth]{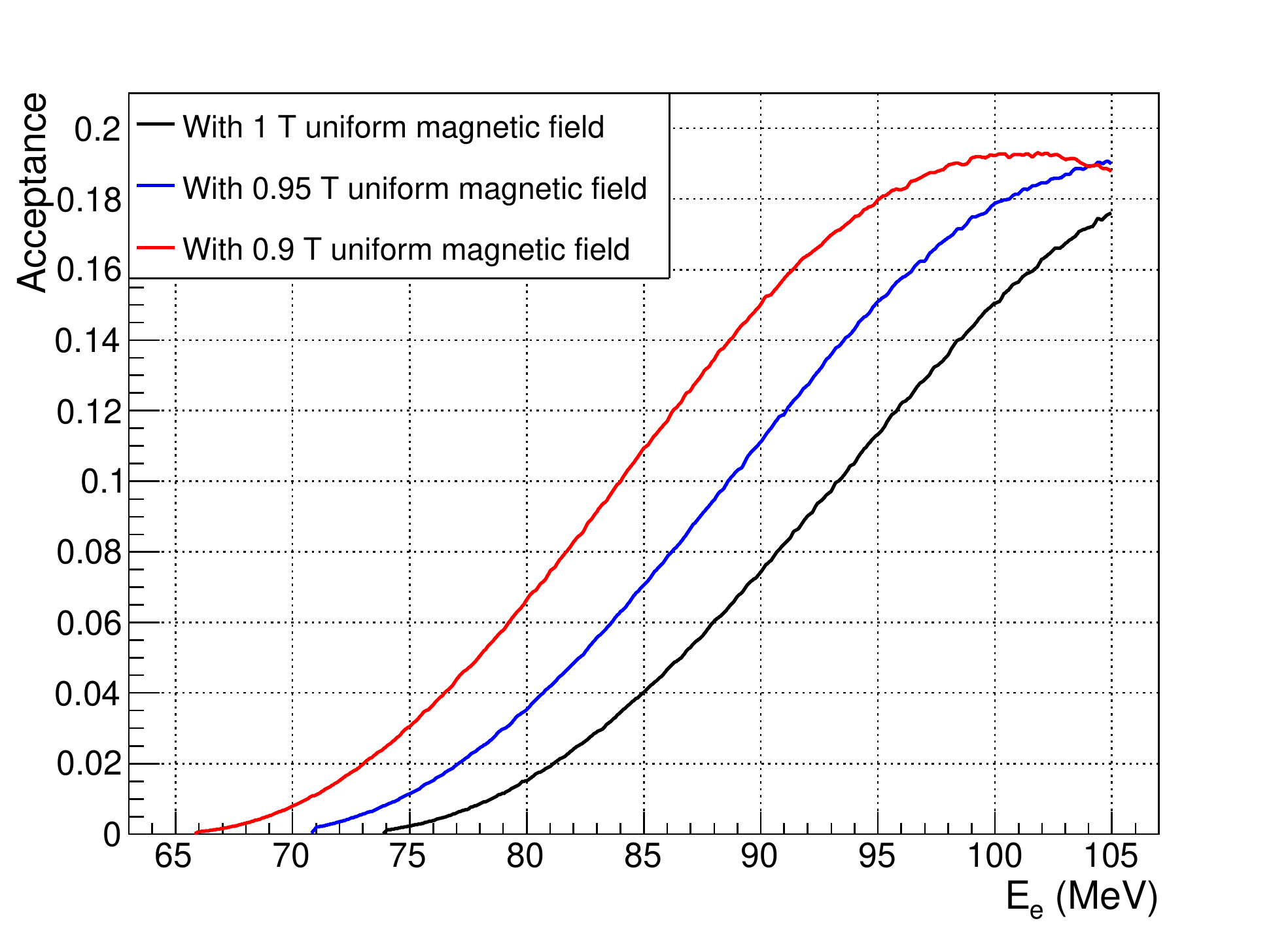}
	\caption{Acceptance distributions of COMET Phase-I by full simulation with uniform magnetic fields of 0.9~T (red), 0.95~T (blue) and 1.0~T (black). The strength of magnetic field is 1.0~T in the official design of COMET Phase-I.}
	\label{fig:acceptancewithEMs}
\end{figure}

To get a credible signal acceptance, a full simulation has been finished utilizing the geometry of detectors described in the COMET technical design report (TDR)~\cite{COMETTDR}. A uniform magnetic field is used within the detector region because the real field map is not available in this work. By scanning the energy of electrons from $74~{\rm MeV}$ to $105~{\rm MeV}$ with step of $25~{\rm keV}$, the acceptance distribution of Phase-I is drawn in Figure~\ref{fig:acceptancewithEMs}, which is the foundation of our study of searching for $\mu \to eJ$ on COMET Phase-I.

As described in Chapter~\ref{chap:3a}, only tracks at least able to reach the inner wall of CyDet will be detected and reconstructed. In this case tracks with low momenta can never be detected if magnetic field exists, which explains that acceptance goes to zero as the electron energy gets lower. Natually it's the most simple and efficiency way to change the acceptance by changing the strength of the magnetic field in the detector area, which impacts the radius of electron tracks.

\subsection{Result of Likelihood analysis}

Finally the expected energy spectrum of electrons can be got by convolving the theoretical spectrum shown in Figure~\ref{fig:spectra} with the acceptance distribution shown in Figure~\ref{fig:acceptancewithEMs}, where the momentum resolution of detectors has been taken into account by smearing the spectrum using a Gaussian distribution with $\sigma = 150~{\rm keV}/c$. Analyzed with the likelihood method in the case of 1.0~T using parameters listed in Table~\ref{tab:paramsPhaseI}, the upper limit of the branching fraction is determined to be
\begin{equation}
{\mathcal B}(\mu \to e J) < 2.3 \times 10^{-5}\ .
\end{equation}

\begin{table}[h]
	\centering
	\renewcommand\arraystretch{1.5}
       \begin{tabular}{|m{1.5cm}<{\centering}|m{3.5cm}<{\centering}|m{3cm}<{\centering}|}
		\hline 
		Parameter      & Description                                               & Value \\ 
		\hline 
		$B_{detector}$ & The strength of the magnetic field in the detector area   & 1 Tesla\\ 
		\hline
		$\sigma_{rec}$ & The momentum resolution of reconstructed tracks           & $150~{\rm keV}/c$ \\ 
		\hline 
		$S.E.S.$       & The single event sensitivity of the COMET experiment      & $3 \times 10^{-15}$ for COMET Phase-I \\
		\hline
		$Acc_{\mu e}$  & The acceptance of $\mu-e$ conversion events               & About 20\% \\
		\hline
		$\Gamma_{{\rm decay}}$ & The ratio of decayed muons over total orbital muons & 0.39 \\
		\hline
		$\Gamma_{{\rm capture}}$ & The ratio of muons captured by nucleus over total orbital muons & 0.61 \\
		\hline
		$E_{signal}$    & The energy range of signal window                       & [74,105]$~{\rm{MeV}}$ \\
		\hline
	\end{tabular}
	\caption{Parameters used in the likelihood analysis of COMET Phase-I}
	\label{tab:paramsPhaseI}
\end{table}

The high event rate of DIO ($R_{\rm DIO}$) could be a limit to the searching for Majoron on the COMET experiment and is concerned about by many people who are interested. Approximately estimated by our analysis, $R_{\rm DIO}$ is given as $16.7~{\rm kHz}$. Compared with the effective trigger rate dedicated by the COMET DAQ system which is not greater than 20 kHz~\cite{COMETTDR}, it seems to be tolerable for the trigger of COMET Phase-I. However it has to be mentioned that there is an optimization of online trigger system performed by COMET Phase-I to reject the tracks of low-energy particles, after which the trigger rate will be reduced to about 1.3 kHz. Therefore some further researches are necessary to achieve a balance between the large quantity of statistics demanded in our research and low event rate required by COMET Phase-I, which are beyond this study.

\subsection{Improve sensitivities by reducing the strength of magnetic field}

As is mentioned above, the significance factor $S$ will get maximum at around $80~{\rm MeV}$, so that a proposal to shift the peak of acceptance distribution to lower energy region by reducing the strength of magnetic field in the detector area has been raised. Two more simulations have been performed with the uniform field of 0.95~T and 0.9~T and the acceptance distributions are got in these two cases, which are shown in Figure~\ref{fig:acceptancewithEMs}. Using the new distributions, the upper limits of ${\mathcal B}(\mu \to e J)$ are calculated at the 90\% C.L. and shown in Table~\ref{tab:resultsWithEMs}.

\begin{table}[h]
\centering
\renewcommand\arraystretch{2}
\begin{tabular}{|c|c|c|c|}
	\hline 
	Magnetic Field & 1T & 0.95T & 0.9T \\ 
	\hline 
	Upper Limit & $2.3 \times 10^{-5}$ & $1.4 \times 10^{-5}$ & $6.9 \times 10^{-6}$ \\ 
	\hline
	Event rate of DIO & $16.7~{\rm kHz}$ & $75.3~{\rm kHz}$ & $394.6~{\rm kHz}$ \\
	\hline
\end{tabular}
\caption{Estimated upper limits after reducing strength of magnetic field of COMET Phase-I}
\label{tab:resultsWithEMs}
\end{table}

Although sensitivity benefits from reducing the strength of magnetic field, the increase of $R_{\rm DIO}$ is unacceptable for the existing design of COMET Phase-I.

\section{Prediction of searching for Majoron on COMET Phase-II}
\subsection{Design of COMET Phase-II}

\begin{figure}[ht]
	\centering
	\includegraphics[width=0.45\textwidth]{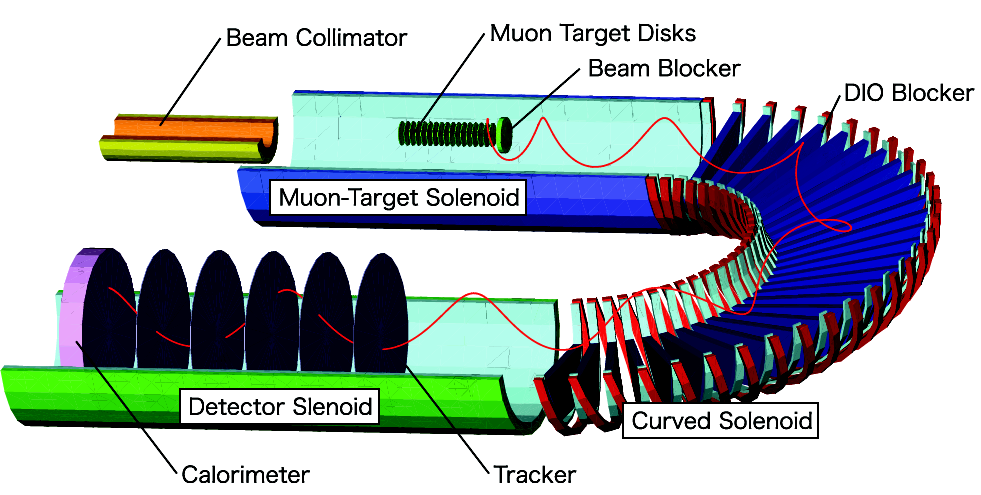}
	\caption{A schematic layout of the solenoid and detectors of COMET Phase-II}
	\label{fig:phaseII}
\end{figure}

COMET Phase-II uses an $8~{\rm GeV}$ proton beam with the electric power of $56~{\rm kW}$ from the J-PARC Main Ring. Pions produced in the collision of proton beam on the target will be captured by a 5~T magnetic field. Inside the transport solenoid bended by $180^{\circ}$ that provides a 3~T magnetic field, the pions will decay into muons. Muons will be stopped by 17 thin flat aluminum disks (muon stopping target) while the electrons will be detected by a spectrometer inside a 1~T magnetic field composed by $20~{\rm \mu m}$ thin straw tubes and an electromagnetic calorimeter (EMC)~\cite{COMETTDR, proceeding}. Figure~\ref{fig:phaseII} shows the layout of the solenoid and detectors.

\subsection{Fast simulation of COMET Phase-II}
\label{chap:4b}

To study the acceptance distribution of COMET Phase-II, a fast simulation is finished without considering material effects and the uncertainty of magnetic field so that the tracks of electrons will be standard helices. Electrons are generated randomly and isotropically from 17 muon stopping targets. In a curved solenoid, the central axis of this trajectory drifts in the direction perpendicular to the plane of curvature. This drift can be estimated by~\cite{COMETTDR}

\begin{equation}
\label{eq:drift}
D = \frac{1}{qB}(\frac{s}{R})\frac{p}{2}(\cos\theta + \frac{1}{\cos\theta}),
\end{equation}
where $q$ is the signed electric charge of the particle, $B$ is the strength of magnetic field along the center-of-beam axis, $\theta$ is the pitch angle of the track, $s$ and $R$ are the path length along the axis direction and the radius of curvature of the curved solenoid respectively. As is shown in Equation~(\ref{eq:drift}), the drifts of electrons with different energy will be different so that a blocker can be placed near the exit of solenoid to eliminate events with low momenta. Finally the energy distribution of electrons at exit of solenoid, multiplied by a roughly estimated detection efficiency, is used to calculate the acceptance distribution.

\begin{figure}[ht]
	\centering
	\includegraphics[width=0.45\textwidth]{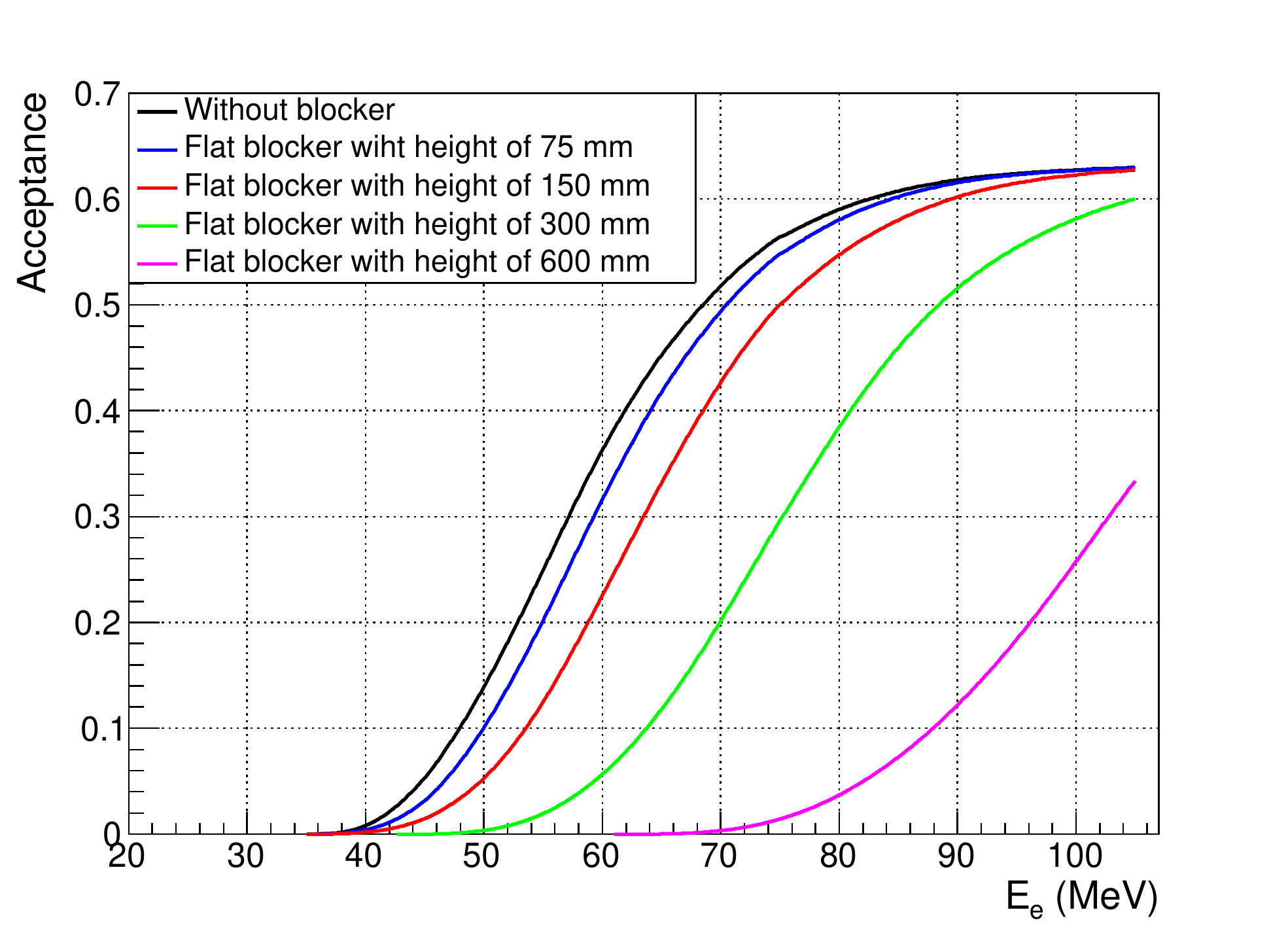}
	\caption{Acceptance distributions got with various heights of flat blocker.}
	\label{fig:acceptancePhaseIIFlat}
\end{figure}

\begin{figure}[ht]
	\centering
	\includegraphics[width=0.45\textwidth]{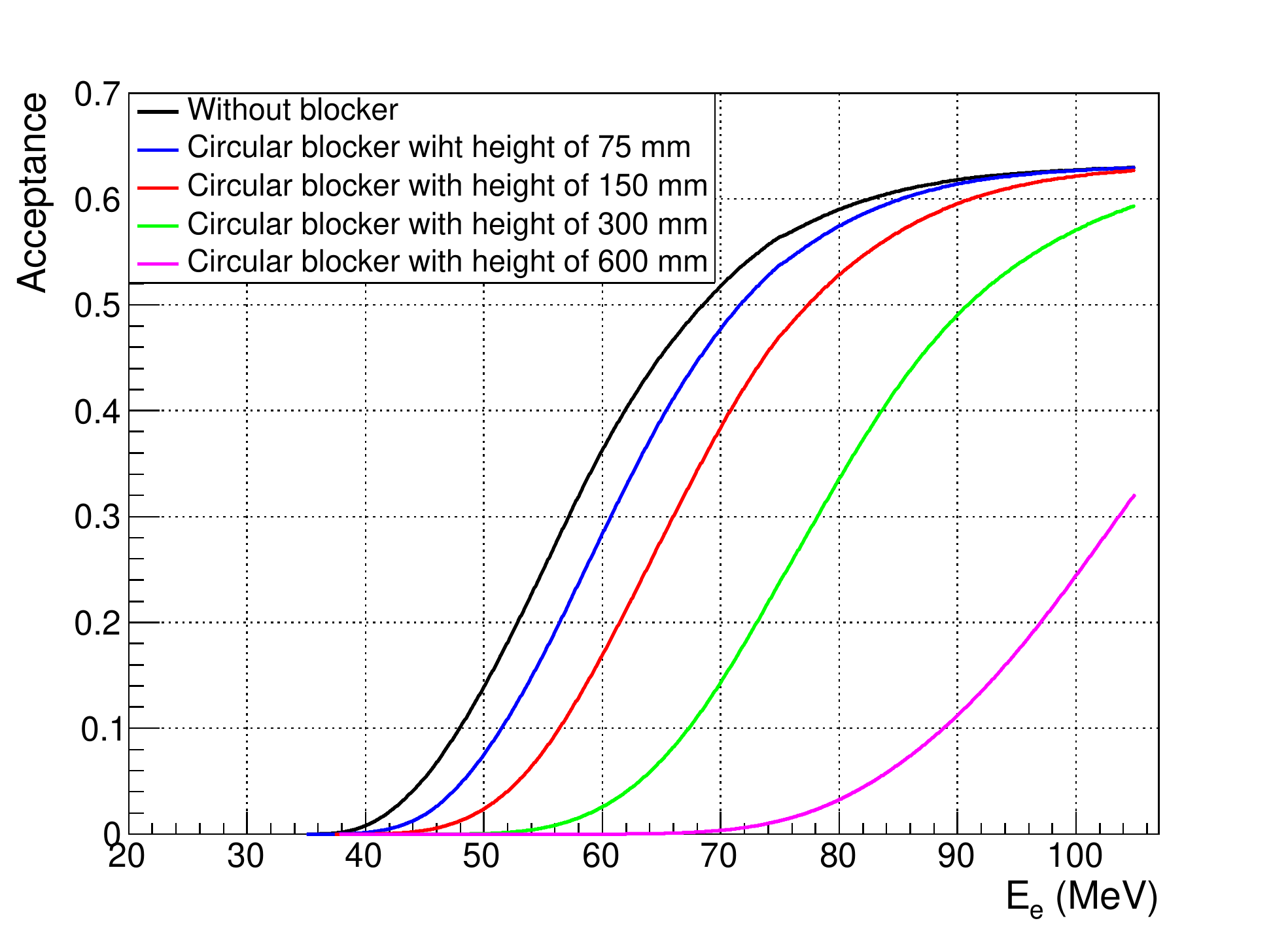}
	\caption{Acceptance distributions got with various heights of circular blocker.}
	\label{fig:acceptancePhaseIICircle2}
\end{figure}

A blocker is designed at the exit of solenoid to control the hit rate in detector, while a large quantity of statistics is demanded in our research to get a more precise spectrum of DIO for comparison. An optimization of blocker is done to balance $R_{\rm DIO}$ and sensitivity. Several designs of blocker are used in this work for comparison since the blocker is still being optimized. In the case of a common design that flat blocker is assembled, the acceptance distributions got by varying the height of the flat blocker from $3~{\rm mm}$ to $600~{\rm mm}$ are shown in Figure~\ref{fig:acceptancePhaseIIFlat}. To optimize the performance, the acceptance distributions got by varying the height of circular blocker with radius of $300~{\rm mm}$ are also used in this study, which are shown in Figure~\ref{fig:acceptancePhaseIICircle2}.

\subsection{Result of likelihood analysis}

With parameters listed in Table~\ref{tab:paramsPhaseII} and $R_{\rm DIO}$ estimated approximately to be $232.4~{\rm MHz}$ by our analysis, the upper limit of searching for $\mu \to eJ$ on COMET Phase-II is estimated using likelihood method to be
\begin{equation}
{\mathcal B}(\mu \to e J) < 4.6 \times 10^{-9},
\end{equation}
where material effects are not considered and blocker is not implemented.

\begin{table}[h]
	\centering
	\renewcommand\arraystretch{1.5}
       \begin{tabular}{|m{1.5cm}<{\centering}|m{3.5cm}<{\centering}|m{3cm}<{\centering}|}
		\hline 
		Parameter      & Description                                                       & Value \\ 
		\hline 
		$B_{detector}$ & The strength of the magnetic field in the detector area               & 1 Tesla\\ 
		\hline
		$\sigma_{rec}$ & The momentum resolution of reconstructed tracks & $150~{\rm keV}/c$ \\ 
		\hline 
		$S.E.S.$       & The single event sensitivity of the COMET experiment            & $2.6 \times 10^{-17}$ for COMET Phase-II \\
		\hline
		$Acc_{\mu e}$  & The acceptance of $\mu-e$ conversion events                       & About 60\% \\
		\hline
		$\Gamma_{{\rm decay}}$ & The ratio of decayed muons over total orbital muons & 0.39 \\
		\hline
		$\Gamma_{{\rm capture}}$ & The ratio of muons captured by nucleus over total orbital muons & 0.61 \\
		\hline
		$E_{signal}$    & The energy range of signal window                               & [35,105]$~{\rm{MeV}}$ \\
		\hline
	\end{tabular}
	\caption{Parameters used in the likelihood analysis of COMET Phase-II}
	\label{tab:paramsPhaseII}
\end{table}

\begin{figure}[ht]
	\centering
	\includegraphics[width=0.45\textwidth]{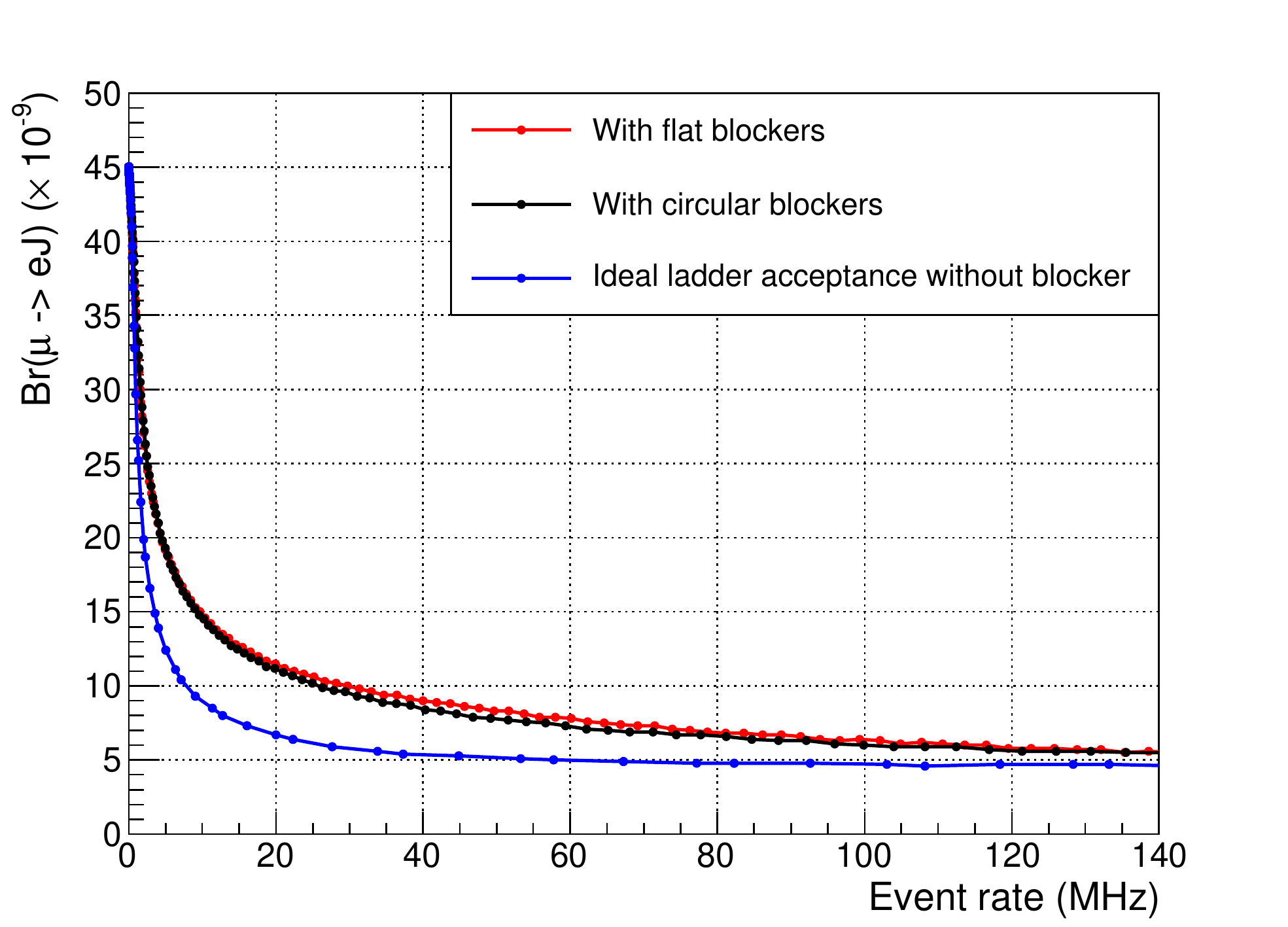}
	\caption{Predicted ${\mathcal B}(\mu \to e J)$ versus $R_{\rm DIO}$ on COMET Phase-II with different designs of blocker ($0~{\rm MHz}<R_{\rm DIO}<140~{\rm MHz}$).}
	\label{fig:PhaseIIResults}
\end{figure}

\begin{figure}[ht]
	\centering
	\includegraphics[width=0.45\textwidth]{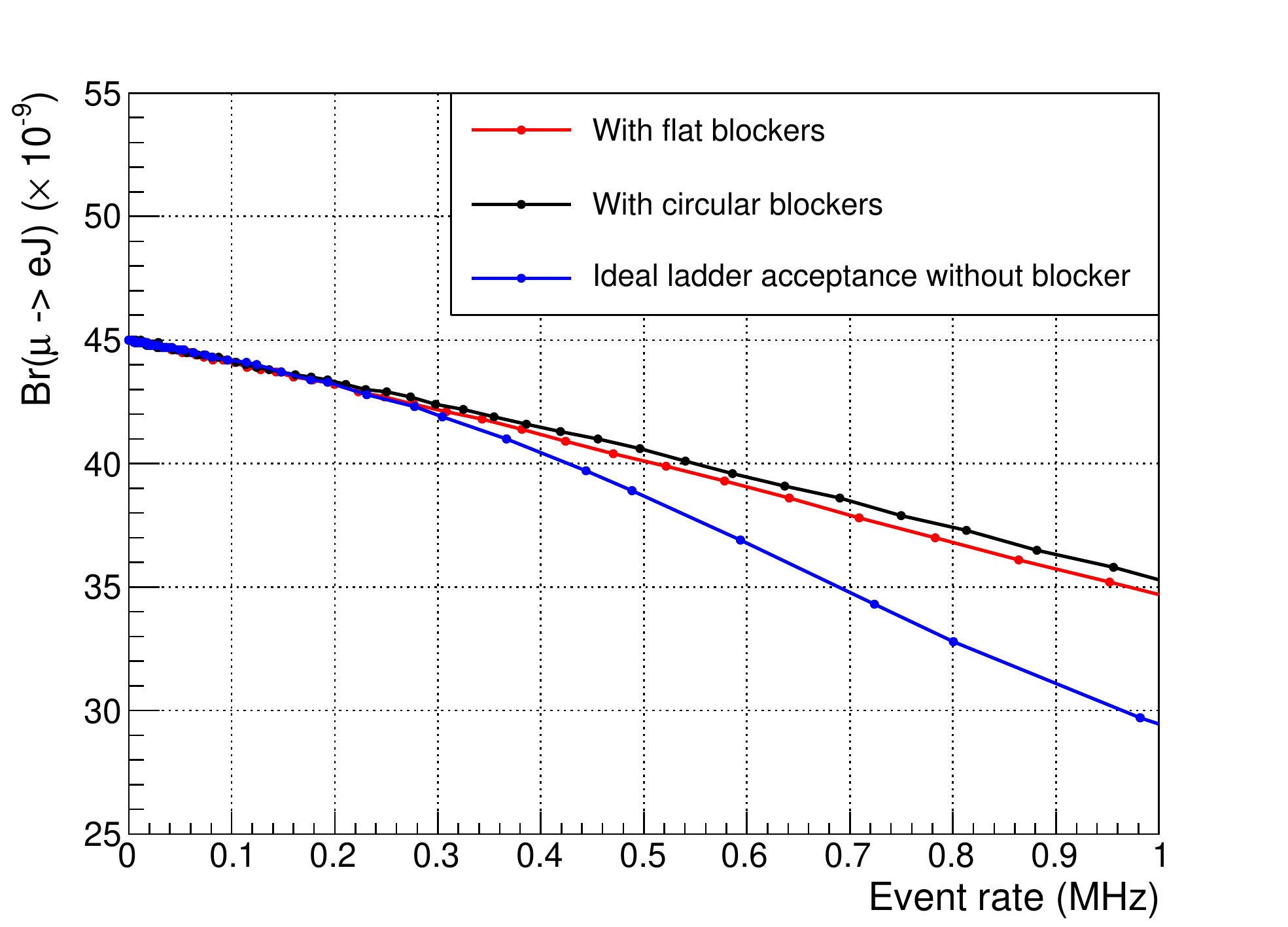}
	\caption{Predicted ${\mathcal B}(\mu \to e J)$ versus $R_{\rm DIO}$ on COMET Phase-II with different designs of blocker ($0~{\rm MHz}<R_{\rm DIO}<1~{\rm MHz}$).}
	\label{fig:PhaseIIResults1}
\end{figure}

Meanwhile, the analysis results with different designs of blocker are shown in Figure.\ref{fig:PhaseIIResults} and Figure.\ref{fig:PhaseIIResults1}. The upper limit would become a little better after simple optimization with circular blockers. Assuming the ideal step-type acceptance distribution with the maximum of 60\%, the relationship between the upper limit and $R_{\rm DIO}$ is calculated when varying the lower limit of signal region, which is used to evaluate the potential of searching for $\mu \to eJ$ on COMET Phase-II.

The likelihood analysis shows that ${\mathcal B}(\mu \to e J)$ varies between $O(10^{-8})$ and $O(10^{-9})$ with different designs of blockers. As is implied in Figure~\ref{fig:PhaseIIResults}, smaller blocker will provide higher precision but also lead to higher $R_{\rm DIO}$, which requires better detector performance. Given that the data will be taken for one year~\cite{proceeding} with the duty factor of $0.2$ (beam-on time divided by the accelerator cycle, which is $0.5/2.48 \approx 0.2$)~\cite{COMETTDR}~\cite{tomizawa}, $R_{\rm DIO}$ varies within the range of $O(10^{2})$ and $O(10^{3})~{\rm MHz}$ if the sensitivity is required to reach $O(10^{-9})$. In this case $R_{\rm DIO}$ is too high for the current detector technology. As is implied in Figure~\ref{fig:PhaseIIResults1}, sensitivity of searching for Majoron on COMET Phase-II can reach $O(10^{-8})$ even with $R_{\rm DIO}$ of $O(10^{2})~{\rm kHz}$, which is a much more realistic prediction.

\section{Discussion of future experiments}
As mentioned in the above sections, according to Equation~\ref{eq:2}, the sensitivity to search for Majoron is directly determined by the yield of muons stopped in the targets and the total number of signal events within detection region which can be estimated by the S.E.S. of $\mu - e$ conversion and the acceptance distribution of detectors, respectively. For future experiments aiming to search for Majoron, it is very important to optimize these two factors according to the real condition of beam and detector. To study the number of stopped moun ($N_{\mu}$) and the number of signal events that can be detected ($N_{detected}$) required to reach a certain sensitivity of the searching for Majoron (${\mathcal B}(\mu \to e J)$), we consider a hypothetical experiment.

\begin{figure}[h]
	\centering
	\includegraphics[width=0.45\textwidth]{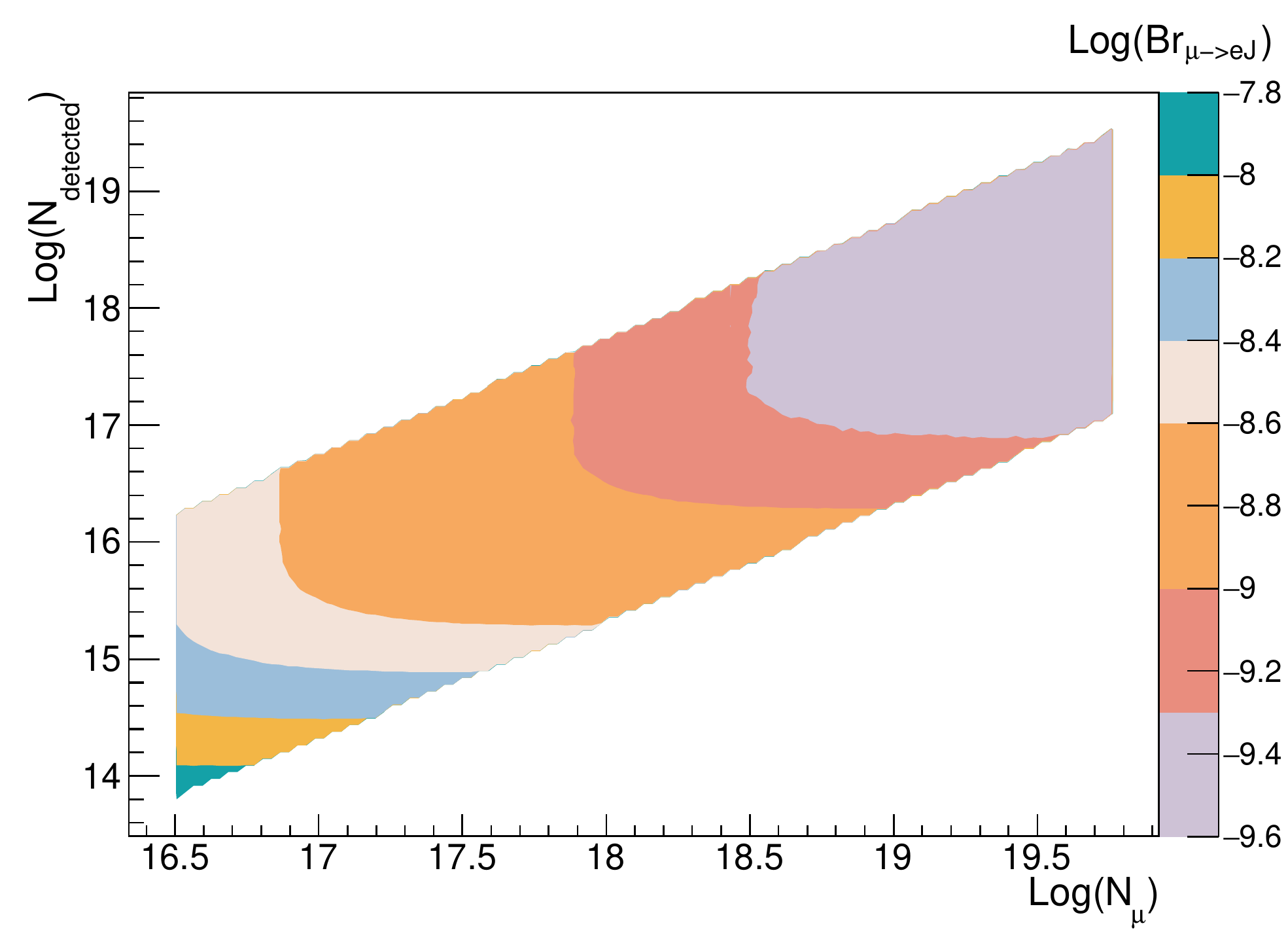}
	\caption{Prediction of searching for Majoron with some ideal assumptions.}
	\label{fig:future}
\end{figure}

Assuming an ideal acceptance of 100\%, without momentum resolution, material effects and backgrounds, the sensitivities to search for Majoron are calculated with various signal windows and S.E.S. of $\mu - e$ conversion, which are shown in Figure~\ref{fig:future}.

To improve the precision, smaller S.E.S. of $\mu - e$ implies that more muons are stopped on the target, which requires more intense beam. Wider signal window means that the energy of detected electrons is closer to the peak of energy spectrum of MEIO, which requires the better performance of detector to tolerate higher event rate. 

It has to be mentioned that there are still other backgrounds rather than DIO from target such as radiative muon capture (RMC), beam background and DIO from the muons captured by the gas in detectors. The measured electron spectrum will be distorted by the contributions from these processes. Since our likelihood analysis is based on the comparison between signal and background spectra, it's no doubt that the sensitivity of searching for Majoron will change after considering other backgrounds. However further researches are necessary to improve the sensitivity due to the lack of the knowledge of these processes, which are beyond the scope of this study.

\section{Conclusion}

As is shown in our study, considering $R_{\rm DIO}$ and current detector technology, the sensitivity to search for Majoron on COMET Phase-I is given as $O(10^{-5})$, which is worse than the current best experimental result. While that will be an improvement of 1000 times on COMET Phase-II in view of the much larger quantity of statistics, which will really be an exciting opportunity to investigate new physics. Furthermore, benefting from more intense beam and better detector, the sensitivity will be significantly improved and is possible to reach the order $O(10^{-10})$.

\section{Acknowledgments}
The authors would like to express thanks to Prof. Yoshitaka Kuno and Dr. Yuichi Uesaka for their strong support, significant effort and precious comments during discussions in Osaka University. The authors are grateful for members of the COMET collaboration for their providing crucial information and valuable help during our work.

This work is supported by the National Natural Science Foundation of China (NSFC) under Contracts Nos. 11335009; International
Partnership Program of Chinese Academy of Sciences, Grant No.113111KYSB20190035.

\bibliography{Tex/references}

\end{document}